\documentclass{article_saj}
\usepackage{graphicx,saj,multicol,subeqnarray}
\usepackage{xcolor}
\usepackage{widetext}
\usepackage{url}
\usepackage{ulem}
\usepackage{bm}
\usepackage{tikz} 
\usepackage{pifont} 
\usepackage{amsfonts}
\usepackage{subfigure}
\usepackage{amssymb}
\usepackage{amsmath,upgreek}
\usepackage{titlesec}
\usepackage{float}
\usepackage{tabularx}
\usepackage[scr=boondoxo,scrscaled=1.05]{mathalfa}

\DeclareUnicodeCharacter{2212}{-}
\graphicspath{ {./Images/} }
\usepackage{biblatex}[style=nature]
\def\point#1{\hbox{\setbox7=\hbox to0.6em{\hfil.\hfil}%
\setbox8=\hbox to0.5em{\hfil$^{#1}$\hfil}%
\box7\kern-0.5em\box8}}

\def\pointmin#1{\hbox{\setbox2=\hbox to0.8em{\hfil.\hfil}%
\setbox3=\hbox to0.6em{\hfil$^{#1}$\hfil}%
\box2\kern-.7em\box3}}

\def\mmm{\pointmin{\mathrm{m}}\kern.15em}

\titlelabel{\thetitle.\quad}
\definecolor{xlinkcolor}{cmyk}{1,0.6,0,0}
\usepackage[bookmarks=false,         
     pdfnewwindow=true,      
     colorlinks=true,    
     linkcolor=xlinkcolor,     
     citecolor=xlinkcolor,     
     filecolor=xlinkcolor,  
     urlcolor=xlinkcolor,      
final=true
]{hyperref}

\begin{document}
\parindent=.5cm
\baselineskip=3.8truemm
\columnsep=.5truecm
\newenvironment{lefteqnarray}{\arraycolsep=0pt\begin{eqnarray}}
{\end{eqnarray}\protect\aftergroup\ignorespaces}
\newenvironment{lefteqnarray*}{\arraycolsep=0pt\begin{eqnarray*}}
{\end{eqnarray*}\protect\aftergroup\ignorespaces}
\newenvironment{leftsubeqnarray}{\arraycolsep=0pt\begin{subeqnarray}}
{\end{subeqnarray}\protect\aftergroup\ignorespaces}
%


\begin{strip}

{\ }

\vskip-1cm

{\ }


\title{Predicting Coupled Electron and Phonon Transport Using the Steepest-Entropy-Ascent Quantum Thermodynamics Framework}


\authors{Jarod Worden$^{a}$, Michael von Spakovsky$^{a}$ and Celine Hin$^{a,b}$}

\vskip3mm

\address{$^a$ME Dept., Center for Energy Systems Research,
Virginia Tech, Blacksburg, VA 24061, United States.}

\address{$^b$MSE Dept., Northern Virginia Center,
Virginia Tech, Falls Church, VA 22043, United States.}


\Email{jaworden@vt.edu}


\begin{abstract}{The principal paradigm for determining the thermoelectric properties of materials is based on the Boltzmann transport equations (BTEs) or Landauer equivalent. These equations depend on the electron and phonon density of states (e-DOS and p-DOS) derived from \textit{ab initio} calculations performed using density functional theory and density functional perturbation theory. Recent computational advances have enabled consideration of phonon-phonon and electron-phonon interactions in these calculations. Leveraging these DOS, the single species BTE or Landauer equivalent can ascertain key thermoelectric properties but overlooks the intrinsic coupling between the e-DOS and p-DOS. To account for this, the multispecies BTE paradigm has, despite its substantial computational burden, been utilized, yielding excellent results in agreement with experiment. To alleviate this computational burden, the steepest-entropy-ascent quantum thermodynamic (SEAQT) equation of motion (EOM), which inherently satisfies both the postulates of quantum mechanics and thermodynamics and predicts the evolution of non-equilibrium states, can be used. Employing the e-DOS and p-DOS as input as well as calculated SEAQT electron and phonon relaxation parameter values that are based on \textit{ab initio} values of relaxation times, group velocities, and effective masses found in the literature, the EOM accurately computes material transport properties, accounting for the e-DOS and p-DOS coupling. It does so at a significantly reduced computational cost across multiple spatial and temporal scales in a single analysis. A succinct overview of the SEAQT framework and its EOM with comparisons of its predictions to measured data for the transport properties of Si, doped Si, and Bi$_2$Te$_3$ is given.}
\end{abstract}


\end{strip}

\tenrm


\section{Introduction}

\indent

Predicting the thermal and electrical transport properties through \textit{ab initio} calculations is crucial to understanding the fundamental physics and thermodynamics governing materials. However, modeling non-equilibrium phenomena, especially in complex material systems with coupled phenomena such as electron-electron, phonon-phonon and electron-phonon coupling poses significant challenges due to the complicated theoretical and heavy computational burdens. Nonetheless, it is precisely these systems that often hold the greatest interest for material applications.

To make such predictions, one needs the densities of states of a material's electrons (e-DOS) and phonons (p-DOS). These can be determined through $ab \; initio$ calculations using density functional theory (DFT)\cite{kresse_efficiency_1996,kresse_ultrasoft_1999,gajdos_linear_2006,hohenberg_inhomogeneous_1964,kohn_self-consistent_1965} and density functional perturbation theory (DFPT) \cite{togo_first_2015,togo_implementation_2023,baroni_phonons_2001}, respectively.  In addition, new codes have emerged enabling the computational assessment of phonon-phonon \cite{li_shengbte_2014,madsen_boltztrap_2006,madsen_boltztrap2_2018,wang_lantrap_2020} and electron-phonon interactions \cite{wang_lantrap_2020,ponce_epw_2016,zhou_perturbo_2021,pizzi_boltzwann_2014} on the e-DOS and p-DOS. Leveraging these density of states, the phonon (lattice) thermal conductivity and electron electrical conductivity can be determined from a complete solution of the single-species Boltzmann transport equation (BTE) or equivalent Landauer formalism in combination with Fourier-based Shankland-Koelling-Wood interpolation \cite{li_shengbte_2014,madsen_boltztrap_2006,madsen_boltztrap2_2018}, Wannier function interpolation \cite{ponce_epw_2016,zhou_perturbo_2021,pizzi_boltzwann_2014,ponce_towards_2018}, or Landauer distribution modes \cite{wang_lantrap_2020}. However, such single-species approaches disregard the impact of the coupling between the electron and phonon densities of states on these transport properties. Incorporating this coupling, which involves the non-equilibrium effect of electron drag on phonons and phonon drag on electrons, is essential for attaining a thorough understanding of the underlying physics and thermodynamics.

To tackle this challenge, Protik and Broido \cite{protik_coupled_2020} successfully developed a fully coupled electron–phonon solution using a set of coupled BTEs to analyze the transport properties of GaAs. Subsequent enhancements by Protik and Kozinsky \cite{protik_electron-phonon_2020}, incorporating fully $ab \; initio$ electron–phonon coupling, were applied to investigate the transport properties of SiC. Further refinements to this methodology were introduced by Protik $et \; al.$ \cite{protik_elphbolt_2022} with their advancements integrated into the $elphbolt$ code assessible through \cite{protik_elphbolt_2021}. In their study \cite{protik_elphbolt_2022}, the authors detail that the computational effort, including determining the e-DOS and p-DOS and various types of electron and phonon scattering and solving the fully coupled electron–phonon BTEs for Si with an n-type doping concentration of 2.75 x 10$^{14}$ cm$^{−3}$ at 300 K, requires approximately 3000 CPU-hr using four nodes each equipped with 28 Intel(R) Xeon(R) CPU E5-2680 v4 $@$ 2.40 GHz cores. The solution domain for the BTEs encompasses 50$\times$50 $\times$50 \textbf{q}-meshes and 150$\times$150$\times$150 \textbf{k}-meshes with a relative convergence threshold set at $10^{-4}$ eV. Comparative analysis against experimental data demonstrates very good agreement \cite{protik_elphbolt_2022}.

An alternative to this Boltzmann or Landauer paradigm, which addresses the complex theoretical and heavy computational burdens, is that provided by the steepest-entropy-ascent quantum thermodynamic (SEAQT) framework \cite{beretta_quantum_1984,beretta_quantum_1985,beretta_nonlinear_2009,beretta_steepest_2014,li_steepest-entropy-ascent_2016,li_steepest_2018,von_spakovsky_trends_2014}. This framework is based on a unification of quantum mechanics and thermodynamics \cite{hatsopoulos_unified_1976,hatsopoulos_unified_1976-1,hatsopoulos_unified_1976-2,hatsopoulos_unified_1976-3} and the development of a nonlinear equation of motion (EOM) \cite{beretta_quantum_1984,beretta_quantum_1985,beretta_nonlinear_2009,beretta_steepest_2014,beretta_general_1981} that extends the time-dependent Schrödinger (or von Neumann) EOM into the realm of irreversible thermodynamic phenomena. The SEAQT EOM, which is based on a gradient dynamic of entropy in state space constrained by the energy, particle number, etc., is guided by the principle of SEA at each instant of time and, therefore, obviates the need for an $a \; priori$ assumption of the kinetic mechanisms involved. This aspect contributes to the generality and versatility of this approach. In fact, the SEA principle has recently been proposed by Beretta as a fourth law of thermodynamics \cite{beretta_fourth_2020}.  In addition, based on the foundational work of Beretta \cite{beretta_quantum_1984,beretta_quantum_1985,beretta_nonlinear_2009,beretta_steepest_2014} and Li and von Spakovsky \cite{beretta_quantum_1984,li_generalized_2016,li_steepest-entropy-ascent_2018}, the SEAQT EOM has been validated and practically applied across multiple spatial and temporal scales to model the non-equilibrium behavior of a variety of quantum and classical systems \cite{cano-andrade_steepest-entropy-ascent_2015,montanez-barrera_loss--entanglement_2020,damian_modeling_2024,montanez-barrera_decoherence_2022,morishita_relaxation_2023,tabakin_local_2023,tabakin_model_2017,ray_no-signaling_2023,ray_steepest_2022,morishita_quantum_2023,mcdonald_predicting_2024,mcdonald_predicting_2023-3,mcdonald_predicting_2023,younis_predicting_2023,mcdonald_entropy-driven_2022,goswami_thermodynamic_2021-1,yamada_kinetic_2020,yamada_low-temperature_2019,yamada_predicting_2019,kusaba_ch4_2019,yamada_method_2018,li_multiscale_2018,kusaba_modeling_2017,li_study_2017,von_spakovsky_predicting_2020}.

By leveraging the e-DOS and p-DOS obtained through $ab \; initio$ calculations for a given material, the SEAQT EOM accounts for the coupled electron-phonon non-equilibrium behavior, enabling predictions of both thermal and electrical transport properties. To scale these results, SEAQT  electron and phonon relaxation parameters are calculated based on \textit{ab initio} values for the eigenenergies, effective masses, group velocities, and relaxation times. Experimental instead of \textit{ab initio} values could also be used. Further details of the theoretical basis for the SEAQT relaxation parameters can be found in \cite{li_steepest_2018} and in Section \ref{Sec2.4} below.

As demonstrated in the following sections, these predictions align well with experimental observations. One key advantage of this method is that the EOM, rooted in the SEA principle and satisfying the postulates of both quantum mechanics and thermodynamics, consists of first-order ordinary differential equations in time, which require minimal computational overhead. Detailed analysis reveals (see Table \ref{TableCPUTimes}) that determining the e-DOS and p-DOS and solving the SEAQT EOM for undoped Si at 300 K takes approximately 36.5 CPU-hr on a single node equipped with an AMD EPYC 7702 chip and 128 cores. Nearly all of this time (98.6\%) is devoted to determining the density of states via DFT and DFPT. Furthermore, predicting the transport properties for 16 different temperatures requires only about 44 CPU-hr since the established DOS need not be recalculated for each temperature.
\begin{table}[h!tbp]
\centering
\caption{CPU times required for obtaining the electron and phonon density of states and running the SEAQT SEAMater code to predict the coupled electron-phonon non-equilibrium behavior of undoped Si. The computations were performed on a single node equipped with an AMD EPYC 7702 chip and 128 cores.}
\smallbreak
\begin{tabularx}{8cm}{|X|X|X|X|} 
  \hline
  {\footnotesize \textbf{e-DOS}} & {\footnotesize \textbf{p-DOS}} \\
  \hline
  {\footnotesize $\sim$32 CPU hr} & 
    {\footnotesize $\sim$4 CPU hr} \\  \hline {\footnotesize \textbf{SEAQT with e-p coupling at 300 K}} & {\footnotesize \textbf{SEAQT with e-p coupling for 16 $\textbf{T}$'s between 250 K and 1000 K}} \\ 
    \hline
    {\footnotesize $\sim$30 CPU min} & 
    {\footnotesize $\sim$8 CPU hr} \\ 
  \hline
\end{tabularx}
\label{TableCPUTimes}
\end{table}

The SEAQT EOM also offers a versatile tool for studying the dynamics of a network of local non-equilibrium systems. This enables the modeling of intricate material structures, including defects and interfaces across various spatial and temporal scales within a single analysis. Coupled with the hypoequilibrium concept \cite{li_steepest-entropy-ascent_2016,li_generalized_2016,li_steepest-entropy-ascent_2018,li_modeling_2016,li_atomistic-level_2014}, the SEAQT framework facilitates the generalization of the Onsager relations and the extension of the thermodynamic equilibrium descriptions of intensive properties (such as temperature, pressure, and chemical potential), the Gibbs relation, etc. to the entire non-equilibrium realm.

The subsequent sections offer a concise introduction to the SEAQT framework and its EOM. This is followed by a description of its implementation in our SEAMater code \cite{von_spakovsky_seamater_2024} and its application to predicting the transport properties of Si and doped Si in time and with temperature. Comprehensive comparisons with experimental data are presented, along with an examination of the outcomes. The discussion concludes with some closing reflections.  

\section{SEAQT Framework}

\indent

The following section details the background of the SEAQT framework, covering the equation of motion and energy eigenstructure, which consists of the the desnity of states of the material, i.e., the discrete eigenenergy spectrum and the degeneracies associated with each of the energy eigenlevels. The development presented here is based on that given in \cite{beretta_steepest_2014,li_steepest-entropy-ascent_2016,li_steepest_2018,hatsopoulos_unified_1976,hatsopoulos_unified_1976-1}.

\subsection{SEAQT Equation of Motion}

\indent

The equation of motion of the SEAQT framework moves through thermodynamic state space (e.g., Hilbert space) where there are $m$ single-particle energy eigenlevels that can be occupied by particles. A system eigenstate is denoted by $|$n$^a$n$^b$...n$^m$〉 where n$^k$ is the occupation number and $\epsilon^k$ the energy of the $k^{th}$ single  particle energy eigenlevel. For fermions, $n^k$ has a value of 0 or 1, and for bosons, a value between 0 and $\infty$. A system thermodynamic state is selected from the Hilbert space spanned by the system eigenstates and can be represented by a $w$-dimensional vector $\{p_{n^k}^k\}$ where $p_{n^k}^k$ represents the probability that $n^k$ particles are observed at the $k^{th}$ single-particle energy eigenlevel. The system thermodynamic state can alternatively be denoted by $\gamma$, which is the square root of the probability vector given by:
\begin{equation}
\gamma = \text{vect}\{\gamma_{n^k}^k\} \equiv \text{vect}(\sqrt{p_{n^k}^k}), \; k = a, ... , m.
\label{eq:1}
\end{equation}
$\gamma$-space is then defined as a manifold $\mathscr{L}$ in Hilbert space whose elements are all of the $w$ vectors of the real finite numbers $X = \text{vect}(x_l)$ and $Y = \text{vect}(y_l)$ with an inner product expressed as
\begin{equation}
(X|Y) = \sum_{l=1}^w x_l y_l.
\label{eq:2}
\end{equation}
Each system property is then defined by a functional $\tilde{A}\left(\gamma\right)$ and its functional derivative by $d\tilde{A}\left(\gamma\right)/d\gamma$.

The time evolution of the thermodynamic state $\gamma\left(t\right)$ obeys the following $w$-vector form of the equation of motion:
\begin{equation}
|d\gamma/dt)= |\Pi_{\gamma})
\label{eq:3}
\end{equation}
where $|\Pi_\gamma)$ is an element of the manifold $\mathscr{L}$ used to describe the local production densities associated with the balance equations for the entropy and the conserved quantities and is derived from the SEA principle subject to a set of conservation laws denoted by $\{\tilde{C}(\gamma)\}$. In this particular development, these include the conservation of energy, $\tilde{H}(\gamma)=\sum_k \sum_n n^k \epsilon_k p_{n^k}^k$ , and of particle number,  $\tilde{N}(\gamma)=\sum_k \sum_n n^k p_{n^k}^k$ , as well as $m$ probability normalization conditions $\tilde{I}_k(\gamma)$. Using Eq. (\ref{eq:2}), the time evolution of the system entropy $\tilde{S}(\gamma)$ and the conserved system properties $\{\tilde{C}(\gamma)\}=\{\tilde{H},\tilde{N},\tilde{I}_a , ... ,\tilde{I}_m\}$ must satisfy the following set of equations:
\begin{equation}
\Pi_{S} \equiv dS/ dt = ( \Phi | \Pi_{\gamma})
\geq 0
\label{eq:4}
\end{equation}
with $| \Phi)
\equiv |\delta\tilde{S}(\gamma)/  \delta\gamma)$ and
\begin{equation}
\Pi_{C_{i}} \equiv dC_{i}/dt = (\Psi_{i} | \Pi_{\gamma})
= 0 
\label{eq:5}
\end{equation}
with $|\Psi_{i})
\equiv | \delta{\tilde{C}}_{i}(\gamma)/ \delta\gamma)$ where $|\Psi_i)$ and $|\Phi)$ are functional derivatives of the system properties defined by Eqs. (A1) to (A4) of \textit{Appendix A} of Li, von Spakovsky, and Hin \cite{li_steepest_2018}.

The time evolution of a system state is a unique trajectory in thermodynamic state space that obeys the SEA principle. This corresponds to a variational problem that finds the instantaneous “direction” of $|\Pi_\gamma)$ that maximizes the rate of entropy production $\Pi_S$ consistent with the conservation constraints $\Pi_{C_i}=0$. This requires the state space to have a metric field with which to determine the norm of $\Pi_\gamma$ and the distance traveled during the evolution of state. The differential of this distance is then expressed as
\begin{equation}
dl = \sqrt{(\Pi_{\gamma} | \hat{G}(\gamma) | \Pi_{\gamma})}dt  
\label{eq:6}
\end{equation}
where $\hat{G}(\gamma)$ is a real, symmetric, positive-definite operator on the manifold  \cite{beretta_steepest_2014,li_steepest-entropy-ascent_2016,li_steepest_2018,li_generalized_2016,li_steepest-entropy-ascent_2018} that is chosen  here to be the identity operator. This choice corresponds to the Fisher-Rao metric. The SEA variational problem then consists of maximizing $\Pi_S$ subject to the conservation constraints $\Pi_{C_i}=0$ and the additional constraint $(dl/dt)^2=\varepsilon$, which ensures that the norm of $\Pi_\gamma$ remains constant. Here, $\varepsilon$ is some small constant. To find the solution, the method of Lagrange multipliers is used such that 
\begin{equation}
\Upsilon = \Pi_{S} - \sum_{i} \beta_{i}\Pi_{C_{i}} - \frac{\tau}{2} ( \Pi_{\gamma} | \hat{G}(\gamma) | \Pi_{(\gamma)} )
\label{eq:7}
\end{equation}
where the $\beta_j$ and $\tau/2$ are the Lagrange multipliers. The variational derivative of $\Upsilon$ with respect to $|\Pi_\gamma)$ is now set equal to zero so that
\begin{equation}
\frac{\delta\Upsilon}{\delta\Pi_{\gamma}} = |\Phi )
- \sum_{i} \beta_{i} |\Psi_{i})
- \tau\hat{G} | \Pi_{\gamma}) = 0\ 
\label{eq:8}
\end{equation}
Solving this last equation for $|\Pi_{(\gamma)})$ yields the SEA equation of motion, i.e.,
\begin{equation}
|\Pi_{\gamma}) = \hat{L}|\Phi)
- \hat{L} \sum_{i} \beta_{i} |\Psi_{i}) 
\label{eq:9}
\end{equation}
In this last expression, $\hat{L}\equiv \hat{G}^{-1}/\tau$ is assumed to be a diagonal operator. The diagonal terms of $\hat{L} \{\tau_{n^k}^k, k=1, ... ,m; n^k = 0,1$ for fermions and $0,1, ... , \infty$ for bosons\} are associated with the relaxation parameters of the system single-particle eigenlevels such that 
\begin{equation}
\hat{L} = \textrm{diag} \Bigl\{\frac{1}{\tau_{n^{k}}^{k}}\Bigl\}
\label{eq:10}
\end{equation}

The values of the Lagrange multipliers are determined by substituting Eq. (\ref{eq:9}) into Eq. (\ref{eq:5}) yielding
\begin{equation}
\sum_{j=1}^{m+2} (\Psi_{i}|\hat{L}|\Psi_{j}) \beta_{j} = (\Psi_{i}|\hat{L}|\Phi)
\label{eq:11}
\end{equation}
where the $\beta_j$ are proportional to the measurements of non-equilibrium system intensive properties such as temperature, pressure, and chemical potential \cite{li_steepest-entropy-ascent_2018}.

Now, substituting the functional derivatives of the system properties as defined in \textit{Appendix A} of Li, von Spakovsky, and Hin \cite{li_steepest_2018}, the equations of motion for the $\gamma^k$ and the $p_{n^k}^k$ of the single-particle energy eigenlevels are expressed as
\begin{eqnarray}
\frac{d\gamma_{n^{k}}^{k}}{dt} &=& \frac{1}{\tau_{n^{k}}^{k}}( - \gamma_{n^{k}}^{k}\text{ln}(p_{n^{k}}^{k}) - n^{k}\epsilon^{k}\gamma_{n^{k}}^{k}\beta_{E} \nonumber \\ &\,& \, -\, n^{k}\gamma_{n^{k}}^{k}\beta_{N} - \gamma_{n^{k}}^{k}\beta_{I}^{k})
\label{eq:12}
\end{eqnarray}
\begin{eqnarray}
\frac{dp_{n^{k}}^{k}}{dt} &=& \frac{1}{\tau_{n^{k}}^{k}}( - 
p_{n^{k}}^{k}\text{ln}(p_{n^{k}}^{k})
- n^{k}\epsilon^{k}p_{n^{k}}^{k}\beta_{E} \nonumber \\ &\,& \,\,-\, n^{k}p_{n^{k}}^{k}\beta_{N} - 
p_{n^{k}}^{k}\beta_{I}^{k})
\label{eq:13}
\end{eqnarray}
where $\beta_E$, $\beta_N$, and $\beta_I^k$ are the Lagrange multipliers corresponding, respectively, to the generators of the motion $\hat{H}$, $\hat{N}$, and $\hat{I}_k$, while $k$ is the single-particle eigenlevel index and $n^k$ the occupation number at that level.

\subsection{Hypoequilibrium concept}

In \cite{li_steepest-entropy-ascent_2016,li_generalized_2016,li_steepest-entropy-ascent_2018}, Li and von Spakovsky introduce the hypoequilibrium concept to simplify the presentation of the equation of motion and facilitate physically interpreting the state evolution. Without significant loss in generality, the assumption is made that the particles in the same single-particle energy $(\epsilon)$ eigenlevel initially are in mutual stable equilibrium relative to the temperature $T^{\epsilon}$ and the chemical potential $\mu^{\epsilon}$. As a result, the initial probability distribution $p_n^{\epsilon}$ of the occupation states is Maxwelliam, i.e.,
\begin{equation}
p_{n^{\epsilon}}^{\epsilon} = e^{- \beta_{I}^{\epsilon} - \beta_{N}^{\epsilon}n^\epsilon - \beta_{E}^{\epsilon}n^{\epsilon}\epsilon} 
\label{eq:14}
\end{equation}
where $\beta_E^\epsilon \equiv \frac{1}{k_B T^\epsilon}$, $\beta_N^\epsilon \equiv \frac{\mu^\epsilon}{k_B T^\epsilon}$, $\beta_I^\epsilon \equiv \text{ln}\Xi^\epsilon$ and
\begin{equation}
\Xi^{\epsilon}( \beta_{E}^{\epsilon},\beta_{N}^{\epsilon} ) = \sum_{n^\epsilon}^{\infty} e^{- \beta_{N}^{\epsilon}n^{\epsilon}}e^{- \beta_{E}^{\epsilon}n^{\epsilon}\epsilon}.
\label{eq:15}
\end{equation}
Such an initial state is an $m^{th}$-order hypoequilibrium state where $m$ is the total number of eigenlevels. 

An additional assumption made is that the different occupation states of the same single-particle eigenlevel have the same relaxation parameter so that
\begin{equation}
\tau_{n^{\epsilon}}^{\epsilon} = \tau^{e} \textrm{ for all } n^{\epsilon} \textrm{with
the same } \epsilon. 
\label{eq:16}
\end{equation}
This makes each relaxation parameter a property of a particular single-particle eigenlevel. 

As is proven in \cite{li_steepest_2018},  with these two assumptions, the equation of motion, Eq. (\ref{eq:13}), maintains the state of the system in a hypoequilibrium state throughout the entire time evolution. As a result, the evolution can be found from the motion of the state of a single-particle eigenlevel using the following equation of motion found by substituting Eq. (\ref{eq:14}) into (\ref{eq:13}): 
\begin{equation}
\frac{dy^{\epsilon}}{dt} = - \frac{1}{\tau^{\epsilon}}( y^{e} - \beta_{E}\epsilon - \beta_{N})
\label{eq:17}
\end{equation}
where
\begin{equation}
y^{\epsilon} = \beta_{N}^{\epsilon} + \beta_{E}^{\epsilon}\epsilon.
\label{eq:18}
\end{equation}

Now, multiplying Eq. (\ref{eq:13}) by the extensive value of the particle number, $\langle N \rangle_\epsilon$, and integrating over $n^\epsilon$ and doing likewise with the extensive values for the energy, $\langle e \rangle_\epsilon$, and the entropy, $\angle s \rangle_\epsilon$, the following equations capture the contributions to these properties from a single-particle energy eigenlevel:
\begin{equation}
\frac{d{\langle N\rangle}_{\epsilon}}{dt} = \frac{1}{\tau^{\epsilon}}A_{NN}^{\epsilon}( y^{\epsilon} - \beta_{E}\epsilon - \beta_{N}) 
\label{eq:19}
\end{equation}
\begin{equation}
\frac{d{\langle e\rangle}_{\epsilon}}{dt} = \epsilon\frac{d{\langle N\rangle}_{\epsilon}}{dt}
\label{eq:20}
\end{equation}
\begin{equation}
\frac{d{\langle s\rangle}_{\epsilon}}{dt} = y^{\epsilon}\frac{d{\langle N\rangle}_{\epsilon}}{dt}
\label{eq:21}
\end{equation}
where $A_{NN}^\epsilon$, which is the particle number fluctuation of the single-particle energy eigenlevel, is given by
\begin{eqnarray}
A_{NN}^{\epsilon} &\equiv& {\langle N^{2}\rangle}^{\epsilon} - ( {\langle N\rangle}^{\epsilon})^{2}  = \frac{\partial^{2}}{\partial^2 \beta_{N}^\epsilon}\textrm{ln}\Xi^{\epsilon} \nonumber \\&=& 
- \frac{\partial{\langle N\rangle}^{\epsilon}}{\partial\beta_{N}^{\epsilon}} = \frac{1}{e^{y} \pm 1}\, \mp \,\frac{1}{( e^{y} \pm 1 )^{2}}.
\label{eq:22}
\end{eqnarray}
Here, fermions are represented with the plus sign and bosons with the negative.

\subsection{Electron transport equation} \label{Sec2.4}

For the transport of electrons between two locations (subsystems) $A$ and $B$, the single-particle energy eigenlevels involved are those of $A$, $\{\epsilon^{A,k}\}$, and $B$, $\{\epsilon^{B,l}\}$. Integrating Eq. (\ref{eq:19}) over the eigenlevels at location $A$ results in 
\begin{eqnarray}
\frac{d{\langle N\rangle}^{A}}{dt} &=& \int \frac{V}{\tau^{A,\epsilon}}A_{NN}^{A,\epsilon}(\beta_{E}^{A,\epsilon}\epsilon + \beta_{N}^{A,\epsilon} - \beta_{E}\epsilon \nonumber \\ &\,& \, - \beta_{N})D^{A}(\epsilon)d\epsilon. 
\label{eq:23}
\end{eqnarray}
In this last expression, $V$ is the volume of subsystem $A$ and $D^A(\epsilon)$ the DOS per unit volume of $A$. In the near-equilibrium region, both subsystems $A$ and $B$ are approximately in stable equilibrium and as a result, $\beta_N^{A,\epsilon}=\beta_N^A$ and $\beta_E^{A,\epsilon}=\beta_E^A$ so that Eq. (\ref{eq:23}) can be rewritten as
\begin{equation}
\begin{split}
\frac{d{\langle N\rangle}^{A}}{dt} = ( \beta_{N}^{A} - \beta_{N}) \int \frac{V}{\tau^{A,\epsilon} }A_{NN}^{A,\epsilon}D^{A}(\epsilon)d\epsilon\, \\
+ ( \beta_{E}^{A} - \beta_{E})\int \frac{V\epsilon}{\tau^{A,\epsilon}}A_{NN}^{A,\epsilon}D^{A}(\epsilon)d\epsilon.
\end{split}
\label{eq:24}
\end{equation}
A similar development holds for $\frac{d⟨N⟩^B}{dt}$.

This near-equilibrium assumption allows use of the zeroth-order approximation of the terms inside the integrals, leaving only the first-order approximation of $d⟨N⟩^A/dt$. Thus, it is assumed that $A$ and $B$ have the same energy eigenstructure and DOS so that 
\begin{equation}
\tau^{A,\epsilon} = \tau^{B,\epsilon} = \tau^{\epsilon} 
\label{eq:25}
\end{equation}
\begin{equation}
D^{A}(\epsilon) = D^{B}(\epsilon) = D(\epsilon)
\label{eq:26}
\end{equation}
and the fluctuations of subsystems $A$ and $B$ are approximately equal to their mutual equilibrium value at $(\beta_N,\beta_E)$, i.e.,
\begin{equation}
A_{NN}^{A,\epsilon} = A_{NN}^{B,\epsilon} = A_{NN}^{\epsilon}\left( \beta_{N},\beta_{E} \right).
\label{eq:27}
\end{equation}

Now, using these last three conditions and particle conservation, the flow from $B$ to $A$ is given by
\begingroup
\thinmuskip=0mu 
\begin{eqnarray}
\frac{d{\langle N\rangle}^{A}}{dt} - \frac{d{\langle N\rangle}^{B}}{dt}  &=& 2\frac{d{\langle N\rangle}^{A}}{dt} \nonumber \\
&=& ( \beta_{N}^{A} - \beta_{N}^{B}) \int \frac{V}{\tau^{\epsilon}}A_{NN}^{\epsilon}D(\epsilon)d\epsilon \nonumber \\
&+& ( \beta_{E}^{A} - \beta_{E}^{B}) \int \frac{V\epsilon}{\tau^{\epsilon}}A_{NN}^{\epsilon}D(\epsilon)d\epsilon.\nonumber \\
\label{eq:28}
\end{eqnarray}
\endgroup
Furthermore, defining $\delta[\beta_E(\epsilon+\mu)] \equiv (\beta_N^A+\epsilon\beta_E^A)-(\beta_N^B+\epsilon \beta_E^B)$ where $\mu \equiv \beta_N/\beta_E$, the total particle flux to $A$, $J_N$, can be written as    
\begin{equation}
J_{N} \equiv \frac{2}{A_c}\frac{d{\langle N\rangle}^{A}}{dt} = 
\frac{1}{A_c}\int \delta\lbrack \beta_{E}(\epsilon + \mu) \rbrack\frac{V}{\tau^{\epsilon}}A_{NN}^{\epsilon}D(\epsilon)d\epsilon
\label{eq:29}
\end{equation}
where $A_c$ is the interface cross-sectional area between subsystems $A$ and $B$. Introducing the following variational relation for every energy eigenlevel:
\begin{equation}
\delta\lbrack \beta_{E}(\epsilon + \mu) \rbrack = \left( (\epsilon + \mu)\frac{d\beta_{E}}{dx} + \beta_{E}\frac{d\mu}{dx}\right)\delta x
\label{eq:30}
\end{equation}
where $\delta x$ is the distance between locations $A$ and $B$, the total flux can be reformulated as
\begin{equation}
J_{N} = \frac{\delta x}{A_c} \int \frac{V}{\tau^{\epsilon}}\left( (\epsilon + \mu)\frac{d\beta_{E}}{dx} + \beta_{E}\frac{d\mu}{dx} \right)A_{NN}^{\epsilon}D(\epsilon)d\epsilon
\label{eq:31}
\end{equation}

If initially the system is in a hypoequilibrium state, the fluctuation $A_{NN}^\epsilon$ can be related to the Fermi distribution $f$ by
\begin{equation}
A_{NN}^{\epsilon} = \beta_{E}^{- 1}\frac{\partial f}{\partial\epsilon}, \;\;
\textrm{ with } f = \frac{1}{e^{\beta_{E}\epsilon + \beta_{N}} + 1} 
\label{eq:32}
\end{equation}
and as a result, the particle flux becomes
\begin{eqnarray}
J_{N} &=& - \frac{V\delta x}{A_c}\left( \frac{dE_{f}^{0}}{dx} + e\mathcal{E} \right)\int\frac{1}{\tau^{\epsilon}}\frac{\partial f}{\partial\epsilon}D(\epsilon)d\epsilon \nonumber \\ 
&+& \frac{V\delta x}{A_c}\beta_{E}^{- 1}\frac{d\beta_{E}}{dx}\int\frac{1}{\tau^{\epsilon}}\left( \epsilon - E_{f} \right)\frac{\partial f}{\partial\epsilon}D(\epsilon)d\epsilon \nonumber  \\
\label{eq:33}
\end{eqnarray}
where $E_f=E_f^0+e\Phi=-\mu=-\beta_N/\beta_E$, $e$ is the electric charge, $\Phi$ the electric field potential, $E_f^0$ the fermi level without an external field, $-d\mu=dE_f^0+ed\Phi$ the differential chemical potential, and $d\Phi/dx=\mathcal{E}$ the external field.

As shown in Ref \cite{li_steepest_2018}, Eq. (\ref{eq:33}), which results from the SEAQT equation of motion, recovers the Boltzmann transport equation (BTE) \cite{wang_two-temperature_2012} in the low-field region even though the former operates in state space and the latter in phase space. As a result, the SEAQT and BTE relaxation parameters $\tau^\epsilon$ and $\tau'$, respectively, can be related via the following relation:
\begin{equation}
\tau^{\epsilon} = \frac{\left( \delta x/v_{x} \right)^{2}}{\tau^{\prime}(\epsilon)} = \frac{3m(\delta x)^{2}}{2\epsilon\tau^{\prime}(\epsilon)}. 
\label{eq:34}
\end{equation}
Here, $m$ is the particle mass and $v_x$ a group or particle velocity.

\subsection{Phonon transport equation}

The derivation of the SEAQT equation for phonon transport is similar to that for electron transport except that there is no particle number operator. Thus, the conservation laws only involve the energy and the probabilities such that $\{\tilde{C}(\gamma)\}=\{\tilde{H}_p,\tilde{I}_a, ... , \tilde{I}_m\}$. As was the case for the electrons, the hypoequilibrium concept is used for the initial states so that the time evolution of the energy at location (subsystem) $A$ is given by
\begin{equation}
\frac{d{\langle E\rangle}^{A}}{dt} = \int \frac{V}{\tau^{A,\epsilon}}\epsilon^{2}A_{NN}^{A,\epsilon}\left( \beta_{E}^{A,\epsilon} - \beta_{E} \right)D^{A}(\epsilon)d\epsilon
\label{eq:35}
\end{equation}

Again, as before, in the near-equilibrium region, $\beta_N^{A,\epsilon}=\beta_N^A$ and $\beta_E^{A,\epsilon}=\beta_E^A$ and Eqs. (\ref{eq:25}) to (\ref{eq:27}) hold so that the energy flow from $B$ to $A$ is written as
\begingroup
\thinmuskip=0mu 
\begin{eqnarray}
\frac{d{\langle E\rangle}^{A}}{dt} - \frac{d{\langle E\rangle}^{B}}{dt} &=& \frac{V}{A_c}\left( \beta_{E}^{A} - \beta_{E}^{B} \right)  \int \frac{\epsilon^{2}}{\tau^{\epsilon}}A_{NN}^{\epsilon}D(\epsilon)d\epsilon \nonumber \\
&=& \frac{V\delta x}{A_c}\frac{d\beta_{E}}{dx} \int \frac{\epsilon^{2}}{\tau^{\epsilon}}A_{NN}^{e}D(\epsilon)d\epsilon. \nonumber \\
\label{eq:36}
\end{eqnarray}
\endgroup

\begin{figure*}
\centering
\includegraphics[width=14cm, height=2.5cm]{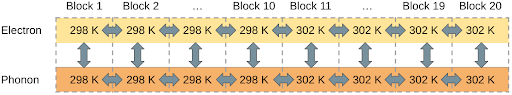}
\caption{Network of local subsystems that show how the local electron and phonon subsystems interact within the SEAQT framework; the arrows indicate electron-electron, phonon-phonon, and electron-phonon interactions.} \label{Fig.1}
\end{figure*}

The fluctuation $A_{NN}^\epsilon$ in this case is based on the boson distribution so that in terms of the Fermi distribution,
\begin{equation}
A_{NN}^{\epsilon} = - \frac{k_{B}T^{2}}{\epsilon}\frac{\partial f}{\partial T}, \;\;
\textrm{ with } f = \frac{1}{e^{\beta_{E}\epsilon} - 1}
\label{eq:37}
\end{equation}
and the energy flux is expressed as
\begingroup
\thinmuskip=0mu 
\begin{eqnarray}
J_{E} &=& \frac{V\delta x}{A_c}\frac{d\beta_{E}}{dx}\int \frac{\epsilon^{2}}{\tau^{\epsilon}}\left( - \frac{k_{B}T^{2}}{\epsilon}\frac{\partial f}{\partial T} \right)D(\epsilon)d\epsilon \nonumber \\ &=& \frac{V\delta x}{A_c}\frac{dT}{dx}\int \frac{\epsilon}{\tau^{\epsilon}}\frac{\partial f}{\partial T}D(\epsilon)d\epsilon  \nonumber \\
&=& \frac{dT}{dx}\int \frac{(\delta x)^{2}}{\tau^{\epsilon}} \hbar \omega \frac{\partial f}{\partial T}D(\omega)d\omega
\label{eq:38}
\end{eqnarray}
\endgroup
where it is noted that for convenience the argument of the integral has been converted from the energy to the frequency $(\omega)$ domain. The relaxation parameter is again chosen based on Eq. (\ref{eq:34}).

\subsection{Electron-Phonon Coupling}

As indicated earlier and shown in \cite{li_steepest_2018}, the BTE in the low-field limit can be recovered from the SEAQT framework via Eq. (\ref{eq:33}). In a similar fashion, the two-temperature model (TTM) of electron-phonon coupling given, for example, in \cite{wang_two-temperature_2012} can be derived from the SEAQT framework as a special case as shown in \cite{li_steepest-entropy-ascent_2018}. Although that development is not presented here, the result is given by the following two equations of motion for $\beta_E^e$ and $\beta_E^p$, which are inversely proportional to the electron and phonon temperatures, respectively:
\begin{equation}
\frac{d\beta_{E}^{e}}{dt} = \frac{{\langle\delta x\rangle}^{2}}{\tau^{e}}\frac{d^{2}\beta_{E}^{e}}{dx^{2}} - \frac{(1 - \chi)}{\tau^{e}}\left( \beta_{E}^{e} - \beta_{E}^{p} \right)
\label{eq:39}
\end{equation}
\begin{equation}
\frac{d\beta_{E}^{p}}{dt} = \frac{(\delta 
x)^{2}}{\tau^{p}}\frac{d^{2}\beta_{E}^{p}}{dx^{2}} - \frac{\chi}{\tau^{p}}\left( \beta_{E}^{p} - \beta_{E}^{e} \right)
\label{eq:40}
\end{equation}
Here $\tau^e$ and $\tau^p$ are the relaxation parameters for the electrons and phonons, respectively, and $\chi$ (see \cite{li_steepest_2018}) is a function of the energy, entropy, and particle fluctuations in the local electron and phonon subsystems that make up a network of such subsystems as shown in Fig. \ref{Fig.1}. The first term on the right of the equals in each expression is the heat diffusion, while the second accounts for the phonon-electron coupling.

Clearly, these equations, as was the case with the BTE, are limited to the near-equilibrium region. To cover the entire non-equilibrium region, even that far from equilibrium, one must return to the equations of motion, Eqs. (\ref{eq:12}) and (\ref{eq:13}), from which the TTM and BTE are derived. It is, in fact, these more general equations, which are used with the network of local subsystems seen in Fig. \ref{Fig.1} to determine the electrical and thermal transport properties of the semiconductor materials modeled here. However, before discussing this, we define in the next sections our transport properties and briefly describe the methods used to obtain our energy eigenstructures.

\subsection{Transport Properties}

The electrical conductivity is defined based on Eq. (\ref{eq:33}) as
\begin{equation}
\sigma = \int \frac{\partial f}{\partial\epsilon}e^{2}\,v_x^{2}\tau^\prime(\epsilon) D(\epsilon)\, d\epsilon
\label{eq:41}
\end{equation}
where the BTE relaxation parameter, $\tau'$, is related to the SEAQT relaxation parameter, $\tau^\epsilon$, via Eq. (\ref{eq:34}), $\delta x^2=V\delta x/A_c$, and the Fermi distribution is that for fermions given in Eq. (\ref{eq:32}).

The thermal conductivity  is expressed as
\begin{equation}
\kappa = \frac{1}{3}\int \tau^{\prime} v_x^{2} C_{\omega}d\omega
\label{eq:42}
\end{equation}
where the specific heat per unit frequency, $C_\omega$, is given by
\begin{equation}
C_{\omega} = \hbar \omega D(\omega)\frac{\partial f}{\partial T}
\label{eq:43}
\end{equation}
and $\hbar$ is Planck’s modified constant. To determine the electron contribution, the Fermi distribution for fermions (Eq. (\ref{eq:32})) is used, while for the phonon contribution, the distribution for bosons (Eq. (\ref{eq:37})) is employed. Note that the temperature $T$ is either the electron or the phonon temperature. 

As to the Seebeck coefficient, it is given  by
\begin{equation}
S = \frac{1}{eT}\frac{\int (\epsilon - \mu)(\partial f/\partial\epsilon)e^{2}\,v_x^{2}\tau^\prime(\epsilon) D(\epsilon)d\epsilon}{\sigma}.
\label{eq:44}
\end{equation}
Here $\mu$ is the chemical potential and $T$ the average of the electron and phonon temperatures.

\onecolumn
\begin{figure}
\centering
\begin{subfigure}
a)
\includegraphics[width=8cm, height=10.5cm]{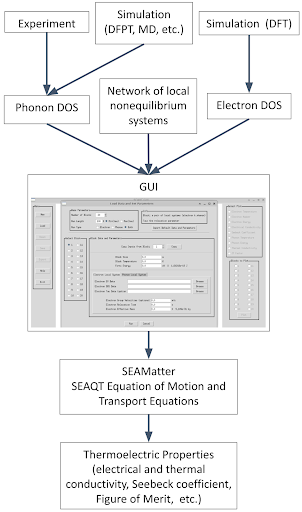}
\end{subfigure}
\begin{subfigure}
\newline
b)
\includegraphics[width=16cm, height=9.5cm]{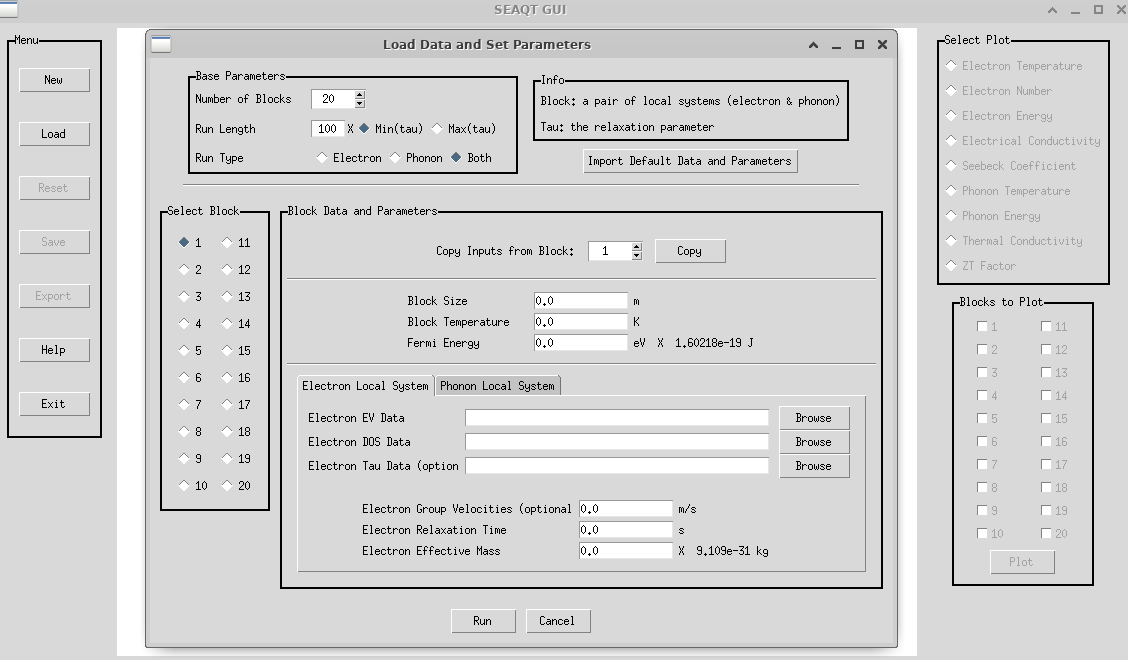}
\end{subfigure}
\caption{a) Workflow diagram of the SEAMatter code for predicting the  thermoelectric properties; b) the GUI showing how the interface displays.} \label{Fig.2}
\end{figure}

\twocolumn
Finally, the figure of merit, $ZT$, which is used to define the performance of semiconductors for thermoelectric applications, is written as
\begin{equation}
ZT = \frac{S^{2}\sigma T}{\kappa}
\label{eq:45}
\end{equation}

\section{SEAQT Implementation and Input Data}

The SEAMatter code, implemented in MATLAB R2021a and Python 3.6, utilizes MATLAB as the primary backend for solving the SEAQT equation of motion with the ODE45 solver \cite{the_mathworks_inc_matlab_2021}. The workflow of the SEAQT concept and SEAMater code are depicted in Fig. \ref{Fig.2}. The energy eigenstructures for both phonons and electrons can be obtained via experimental or simulation methods. In addition, accurate determination of the phonon relaxation parameter and phonon velocity is essential for reliable results.

In this work, the electron energy eigenstructure, represented by the e-DOS, is determined using density functional theory (DFT) with VASP. The Local Density Approximation (LDA) pseudo-potential for cell relaxation and the additional hybrid functional HSE06 for static band structure calculations is employed. The hybrid functional is utilized to account for the common small band gap problem observed in semiconductor materials using DFT methods. The phonon energy eigenstructure is calculated using DFPT along with the post-processing phonon calculation code phonopy \cite{togo_implementation_2023}. The input data for the Si e-DOS and p-DOS along with the relaxation parameters is provided in $Supplementary \; Materials$. 

Additional necessary inputs come from the network of local non-equilibrium subsystems, shown in Fig. \ref{Fig.1}, that describes the number of such subsystems and their respective sizes. These inputs are then fed into the SEAMater code via the GUI after which the thermoelectric property predictions are made. Although the most optimized approach involves utilizing both the e-DOS and p-DOS, calculations using only one type of DOS can still be made, providing inherent material properties specific to either electrons or phonons only.

To compute the total rate of change of energy and mass within a given local subsystem, the energy and mass exchanges with neighboring local subsystems at each instant of time must be determined. For example (see Fig.~\ref{Fig.1}), Electron-2 interfaces with three neighbors: Electron-1, Electron-3, and Phonon-2. Calculating the mass and energy transfers from Electron-1 to Electron-2 involves applying the SEAQT equations of motion, Eqs. (19) and (20), to a composite system comprised of the Electron-1 and Electron-2 subsystems under specific constraints. This composite system constitutes a hypo-equilibrium description of $(m+m')^{th}$-order, where $m$ represents the number of eigenlevels in Electron-1 and $m'$ the number in Electron-2. The mass and energy transfers result from the relaxation of this non-equilibrium composite system. Electron-electron transport is resolved subject to the constraints on $\{\tilde{C}(\gamma)\}=\{\tilde{H}_e,\tilde{N},\tilde{I}_a, ... , \tilde{I}_m\}$, phonon-phonon transport to the constraints on $\{\tilde{C}(\gamma\}=\{\tilde{H}_p,\tilde{I}_a, ... , \tilde{I}_m\}$, and phonon-electron transport to the constraints on $\{\tilde{C}(\gamma)\}=\{\tilde{H},\tilde{N},\tilde{I}_a, ... , \underline{I}_m\}$. In the latter case, $\tilde{H}$ represents the contributions from both the electrons and phonons. This process is repeated for other pairings with Electron-3 and Phonon-2, resulting in the total rate of energy and mass change of Electron-2. By employing this method, the energy and mass exchanges in all local subsystems are determined.

\section{Results}

Results are presented for the electrical and thermal conductivities, the Seebeck coefficient, and the figure of merit for both Si and doped Si. The calculated SEAQT electron and phonon relaxation parameter values used for these results are provided in \textit{Supplementary Materials}. In addition, supplementary materials demonstrate SEAQT's ability to predict the properties of anisotropic materials, as evidenced by the results obtained for Bi$_2$Te$_3$ given in the \textit{Supplementary Materials}.

\subsection{Semiconductor Investigation}

Figure 3 demonstrates that the SEAQT predictions closely match the values and trends observed in various experimental studies \cite{dabbadie_enhancement_2013,fulkerson_thermal_1968,shanks_thermal_1963} for the electrical and thermal conductivities, Seebeck coefficient, and figure of merit of Si. While there are some discrepancies, particularly in thermal conductivity and $ZT$ factor, these differences potentially stem from crystal defects inherent in the fabrication process, affecting phonon velocities and lifetimes and, consequently, reducing the thermal conductivity. Since the SEAQT simulations are based on a perfect Si crystal structure, the calculated thermal conductivity tends to be higher than the experimental values. However, such point and extended defects can be accommodated by the SEAQT framework by obtaining the e-DOS and p-DOS and modified relaxation parameters and group velocities of the specific materials with said defects via experimental methods or DFT, DFPT or molecular dynamics. This will be described in a followup paper.

\begin{figure}[!ht]
\centering
         a) 
    \begin{minipage} {0.45\textwidth}
        \centering \includegraphics[width=7.5cm, height=5.06cm]{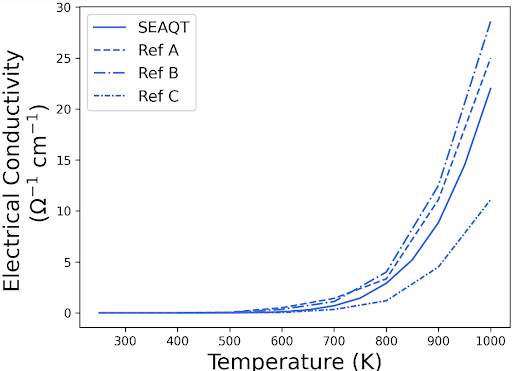} 
    \end{minipage}
        b) 
    \begin{minipage}{0.45\textwidth}
        \centering \includegraphics[width=7.5cm, height=5.06cm]{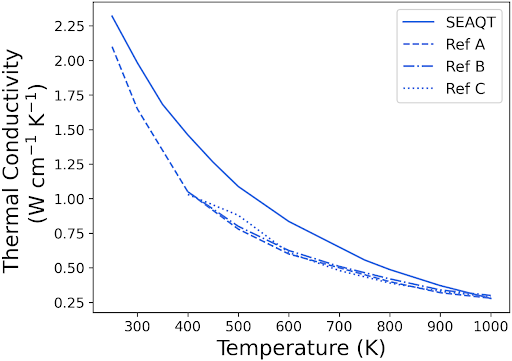} 
     \end{minipage} 
\centering
        c)
    \begin{minipage}{0.45\textwidth}
        \centering\includegraphics[width=7.5cm, height=5.06cm]{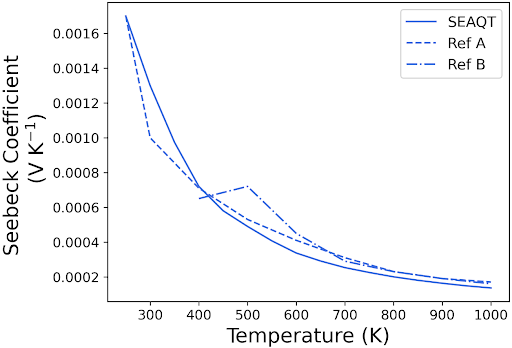} 
    \end{minipage}\hfill
        d)
    \begin{minipage}{0.45\textwidth}
        \centering
        \includegraphics[width=7.5cm, height=5.06cm]{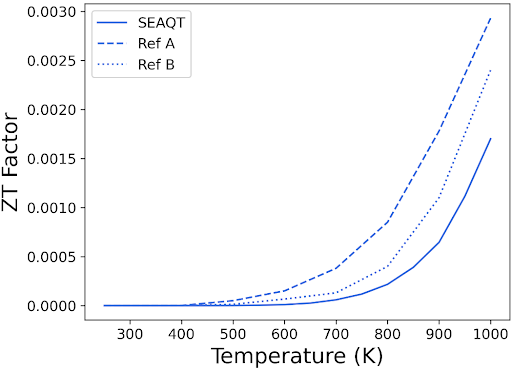} 
    \end{minipage}
\caption{a) Comparison of the SEAQT results for the electrical conductivity, b) thermal conductivity, c) Seebeck coefficient, and d) ZT factor of Si with experimental data: Ref A is \cite{dabbadie_enhancement_2013}, Ref. B is \cite{fulkerson_thermal_1968}, and Ref. C is \cite{shanks_thermal_1963}.} \label{Fig.3}
\end{figure}

\subsection{Doped Si}

Extrinsic Si semiconductors are investigated within the SEAQT framework with either Boron (p-type) or Phosphorous (n-type) doping elements. Doping can also emanate from charged vacancies, although their concentration typically has minimal impact on semiconductor electrical properties compared to doping elements \cite{hoover_nonequilibrium_1992}. Doping elements are employed to boost electrical conductivity at lower temperatures, which is achieved either by augmenting the electron concentration in the conduction band (n-type doping), leading to a rise in the Fermi level, or by introducing holes in the valence band (p-type doping), resulting in a decrease in the Fermi level \cite{hoover_nonequilibrium_1992}. It is posited here that Si doping levels insignificantly affect the electron DOS but can be taken into account by altering the Fermi level, a crucial step in modeling the modified transport properties of doped Si within the SEAQT framework. This is done using the techniques outlined in Refs. \cite{madsen_boltztrap2_2018,smets_solar_2016}. The chosen doping levels are informed by experimental n- and p-type doping levels for comparative analysis with predicted outcomes \cite{pearson_electrical_1949}. The principal alteration in phonon transport properties arises from variations in phonon relaxation due to additional scattering by impurities, free electrons/holes, and bound electrons/holes. This can be taken into account without substantial adjustments to the phonon DOS \cite{asheghi_thermal_2002}. Calculated phonon relaxation parameters are tabulated and provided in the \textit{Supplementary Materials} in Table S1 \cite{asheghi_thermal_2002}.

\begin{figure*}[!ht]
\centering
        a) 
    \begin{minipage}{0.45\textwidth}
        \centering
        \includegraphics[width=8cm, height=5.5cm]{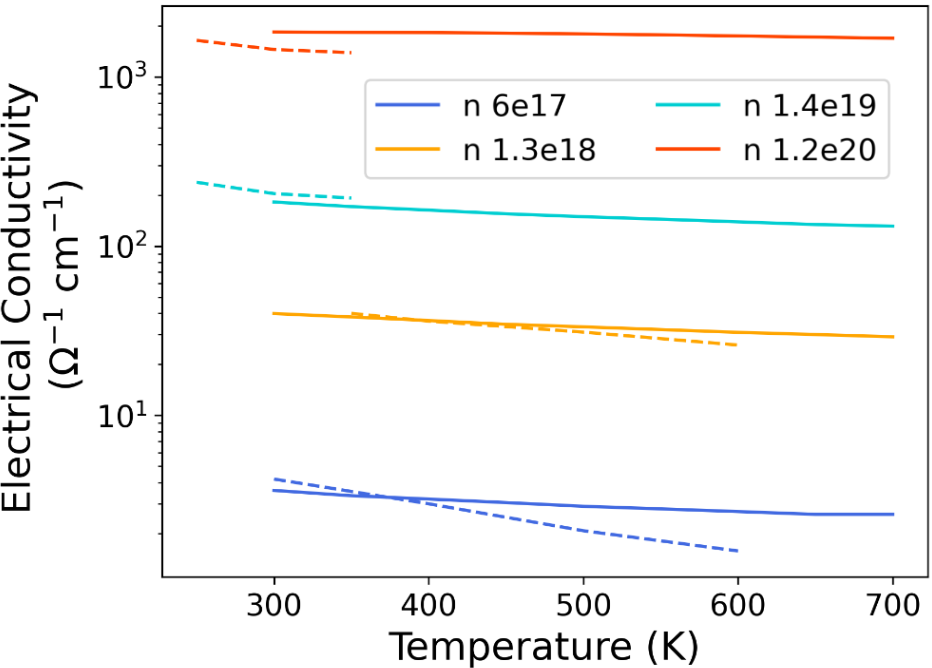} 
    \end{minipage}\hfill
        b) 
    \begin{minipage}{0.45\textwidth}
        \centering
        \includegraphics[width=8cm, height=5.5cm]{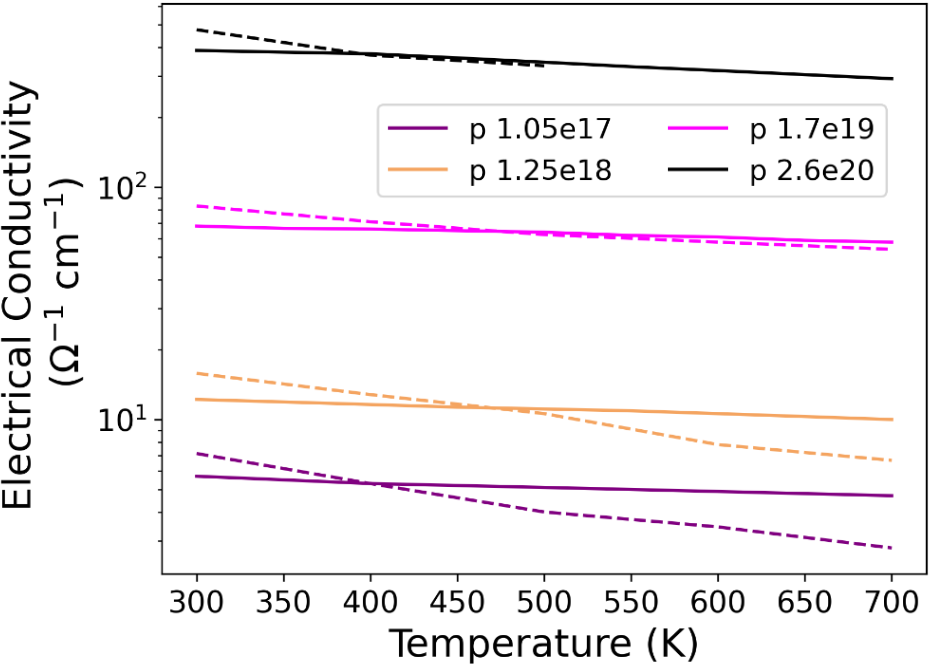} 
    \end{minipage}
\centering

    c) 
    \begin{minipage}{0.45\textwidth}
        \centering
        \includegraphics[width=8cm, height=5.5cm]{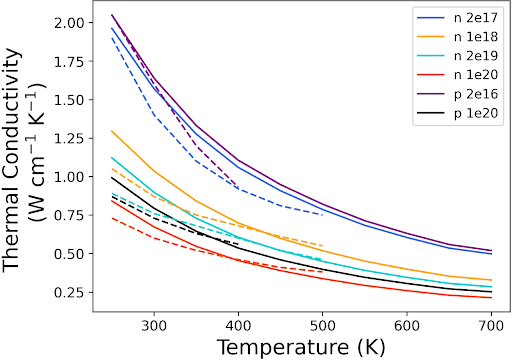} 
    \end{minipage}\hfill
        d) 
    \begin{minipage}{0.45\textwidth}
        \centering
        \includegraphics[width=8cm, height=5.5cm]{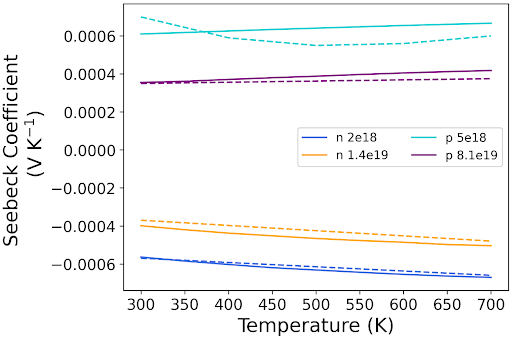} 
    \end{minipage}
\centering

        e) 
    \begin{minipage}{0.45\textwidth}
        \centering
        \includegraphics[width=8cm, height=5.5cm]{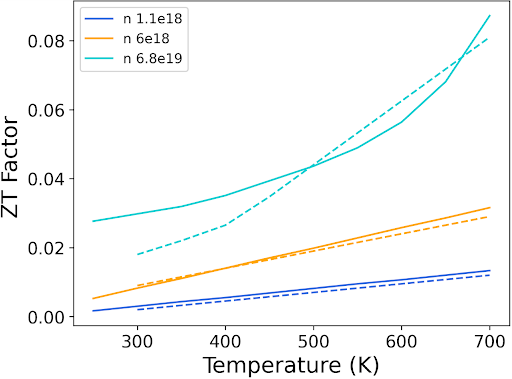} 
    \end{minipage}\hfill
    f)
    \begin{minipage}{0.45\textwidth}
        \centering
        \includegraphics[width=8cm, height=5.5cm]{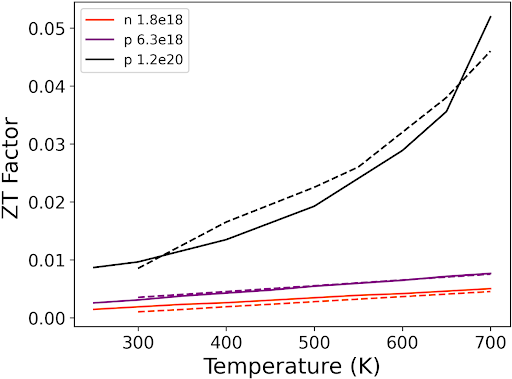} 
    \end{minipage}
\caption{a) and b) Comparisons of the SEAQT results (solid curves) for the electrical conductivity \cite{pearson_electrical_1949}, c) thermal conductivity \cite{asheghi_thermal_2002}, d) Seebeck coefficient \cite{stranz_thermoelectric_2013}, and e) and f) $ZT$ factor \cite{ohishi_thermoelectric_2015} of doped Si with experimental data (dashed curves). } \label{Fig.4}
\end{figure*}

As depicted in Figs. 4a), b), and d), the predicted SEAQT values for the electrical conductivity and Seebeck coefficient align well with experimental results, demonstrating an increase in electrical conductivity with increased doping, a decrease in the absolute Seebeck coefficient with increased doping, and the expected trends for increasing temperature. SEAQT predictions for thermal conductivity, encompassing contributions from both electrons and phonons, are illustrated in Fig. 4c. The SEAQT results for n-doping with Phosphorus and p-doping with Boron closely mirror experimental values, with the most notable deviation occurring for the case of the smallest n-doping. The thermal conductivity for the doped Si consists of a phonon contribution, $\kappa_l$ and an electron contribution, $\kappa_e$ such that the total thermal conductivity $\kappa = \kappa_l + \kappa_e$. $\kappa_e$ is proportional to the electrical conductivity, $\sigma$, via the Lorentz number such that
\begin{equation}
\frac{\kappa_{e}}{\sigma} = LT 
\label{eq:46}
\end{equation}
where \textit{T} is the temperature and \textit{L} the Lorentz number that ranges approximately from 2.4×10$^{-8}$ $W\Omega K^2$ for degenerate semiconductors to 1.5×10$^{-8}$ $W\Omega K^2$ for non-degenerate semiconductors \cite{kim_characterization_2015,thesberg_lorenz_2017}. 

\begin{figure} [!ht]
\centering
\includegraphics[width=8cm, height=5.5cm]{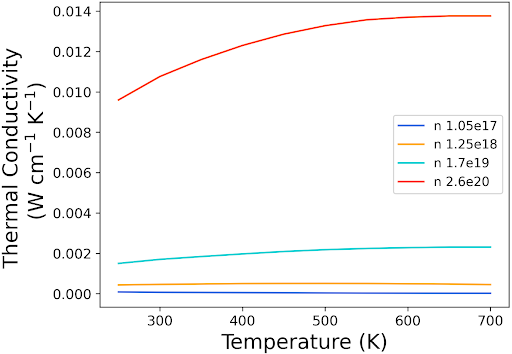}
\caption{Electron thermal conductivity of doped Si.} \label{Fig.5}
\end{figure}

As is evident from this last equation, at lower electrical conductivities, the electron contribution to the total thermal conductivity remains minimal but escalates with higher doping levels corresponding to increased electrical conductivity. Fig. \ref{Fig.5} illustrates that while higher doping leads to enhanced electrical conductivity, the increment is insufficient to significantly influence the total thermal conductivity of doped Si.

Finally, there are minor disparities between the thermal conductivity results obtained from SEAQT and the experimental data, with the SEAQT values being slightly higher. This could potentially stem from adjustments in the relaxation parameter or the exclusion of other defects such as dislocations in the experimental samples, which would impact the thermal conductivity. Despite these slight differences, the SEAQT \textit{ZT} factor closely matches experimental data across all doping levels. Moreover, the electron contribution to the total thermal conductivity of a conventional semiconductor like Si is small. This is not the case for narrow-bandgap semiconductors like Bi$_2$Te$_3$. The SEAQT results obtained for Bi$_2$Te$_3$ given in the \textit{Supplementary Materials} clearly demonstrate this.

\section{Conclusions}
The SEAQT framework is an effective approach for accurately predicting the transport properties of semiconductor materials. SEAQT takes into account electron and phonon energy eigenstructures to determine material transport properties. Furthermore, this framework inherently satisfies the laws of quantum mechanics and thermodynamics, utilizing an equation of motion to determine unique non-equilibrium thermodynamic paths through Hilbert space and is able to cross several spatial and temporal scales in a single analysis. 

The SEAQT framework, which has been used to accurately determine the transport properties of Si and doped Si (as well as Bi$_2$Te$_3$ in the \textit{Supplementary Materials}), provides results similar to those found with current computational methods but does so with a significantly reduced computational burden. The robustness of the SEAQT framework is such that it can be applied to a multitude of systems and can be used to analyze the effect of defect structures and thermoelectric breakdown. This is left for future work.

\section*{Acknowledgements}
The authors would like to thank Virginia Tech's Advanced Research Computing Center (ARC) for the use of its high-performance supercomputing clusters. The lead author would also like to thank the Nuclear Regulatory Commission (NRC) for its  financial support (grant number 31310019M0045) and the support with internal funds provided by the second author and by the Virginia Tech College of Engineering.  

\section*{Data Availability}
The underlying code for this study is not publicly available but may be made available to qualified researchers on reasonable request from the corresponding author.

\printbibliography

\end{document}


\date{}

\maketitle

\section*{Appendix A: Si and doped Si}
\smallbreak

Figs. S1 a) and b) depict the electron and phonon density of states, respectively. Utilizing a general electron relaxation parameter value of $10^{-14}$ s found in the literature \cite{Li_study_2009,Ryu_computational_2016,Zhu_significant_2020}, the spectrum 
 of SEAQT electron relaxation parameter values are calculated using the expression after the second equal in Eq. (34) of the main text. This spectrum appears in Figure S2. To calculate the SEAQT phonon relaxation parameter value, the expression after the first equal in Eq. (34) is used as is a value of 6100 m/s for the phonon group velocity obtained from DFPT calculations with Phonopy \cite{togo_implementation_2023} and a value of 14.93 ns for the relaxation time obtained from MD calculations \cite{henry_spectral_2008}. Values for the calculated SEAQT phonon relaxation parameter appear in Table S1.
\begin{figure} \label{FigSiDOS}  
    a)
    \begin{minipage}{0.45\textwidth}
        \centering
        \includegraphics[width=8cm, height=6.5cm]{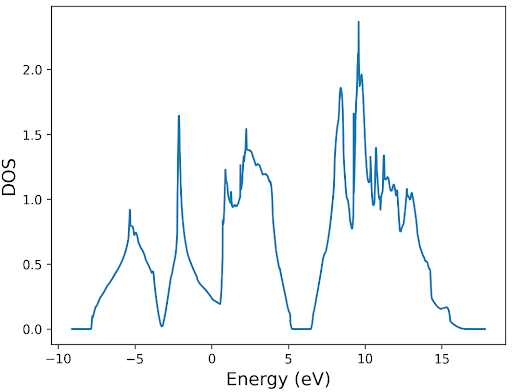} 
    \end{minipage}\hfill
    b)
    \begin{minipage}{0.45\textwidth}
        \centering
        \includegraphics[width=8cm, height=6.5cm]{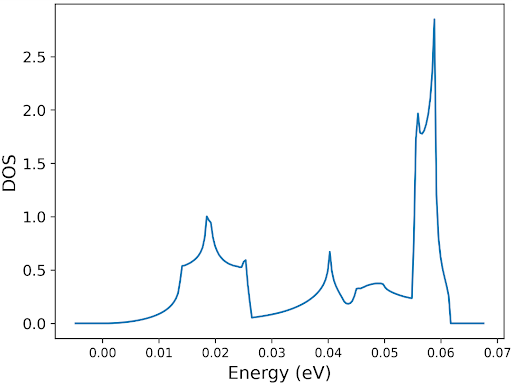} 
    \end{minipage}\hfill
\caption*{Fig. S1 a) Electron density of states and b) phonon density of states for cubic Si.} 
\end{figure}

\begin{figure}
\centering
    \begin{minipage}{0.45\textwidth}
        \centering
        \includegraphics[width=8.5cm, height=6.5cm]{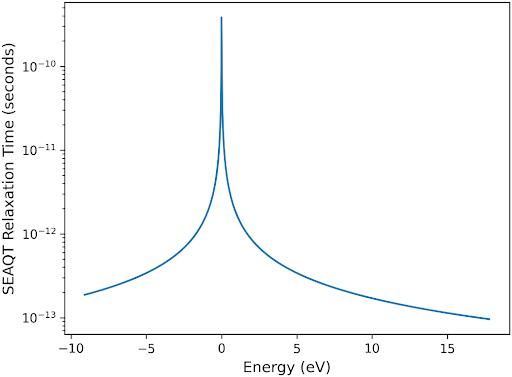} 
    \end{minipage}\hfill
\caption*{Figure S2: Calculated SEAQT electron relaxation parameter ($\tau^\epsilon$) values for Si as a function of the electron eigenenergies. }
\end{figure}

\begin{figure}
\centering
    \begin{minipage}{0.45\textwidth}
        \centering
        \includegraphics[width=8.5cm, height=6.5cm]{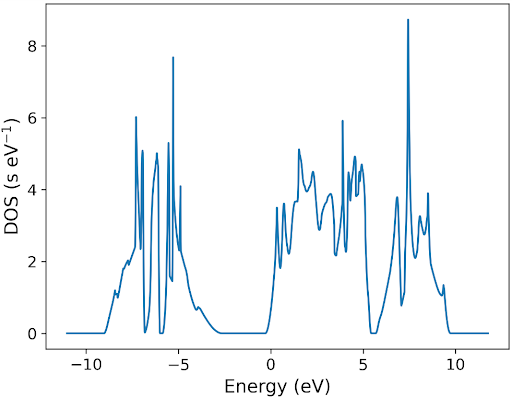} 
    \end{minipage}\hfill
\caption*{Figure S3: Electron DOS of Bi$_2$Te$_3$.}
\end{figure}

\begin{figure}
\centering
    \begin{minipage}{0.45\textwidth}
        \centering
        \includegraphics[width=8.5cm, height=6.5cm]{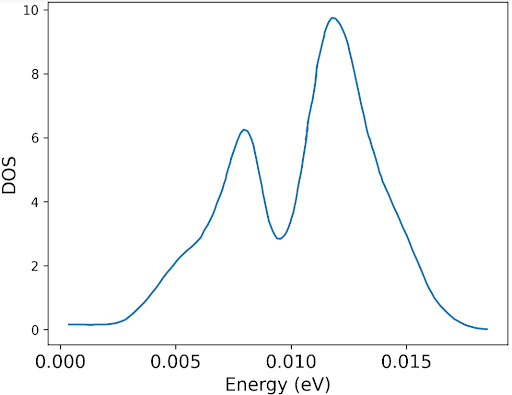} 
    \end{minipage}\hfill
\caption*{Figure S4: Phonon DOS of Bi$_2$Te$_3$ \cite{li_thermal_2015}.}
\end{figure}

\section*{Appendix B: Bi$_2$Te$_3$}
Bi$_2$Te$_3$ serves as a case study for applying the SEAQT framework to predicting the transport properties of a narrow-bandgap semiconductor. The presence of spin-orbit coupling between Bi and Te atoms introduces additional computational intricacy, particularly concerning the electron density of states (DOS) compared to simpler semiconductor systems. VASP is used to determine the e-DOS as shown in Fig. S3. Conversely, the phonon DOS, depicted in Fig. S4, is extracted from first-principle DFPT calculations \cite{li_thermal_2015}. In addition, the SEAQT phonon relaxation parameter ($\tau^\epsilon$) plays an important role in scaling the material's thermal conductivity, particularly in the context of anisotropic materials like Bi$_2$Te$_3$ where the thermal conductivity varies along the $xx$- or $zz$-directions (i.e., transverse and longitudinal directions) \cite{hellman_phonon_2014}. This anisotropy influences phonon group velocities, which are sourced from \cite{alnofiay_brillouin_2014} and presented in Table S2. The SEAQT phonon relaxation parameter ($\tau^\epsilon$) values calculated using the expression after the first equal in Eq. (34) of the main text also appear in Table S2 and are based on the phonon group velocities and a phonon relaxation parameter ($\tau^\prime$) value of 2.210$^{-11}$s obtained from \cite{feng_prediction_2014}. Per Eq. (34) of the main text, the SEAQT phonon $\tau^\epsilon$  does not vary with the eigenenergies since $\tau^\prime$ does not in this case. 

As to the SEAQT electron $\tau^\epsilon$, these are affected by the anisotropic change of the effective mass as well as by the eigenenergies (see the expression after the second equal in Eq. (34) of the main text). These masses are given in Table S3 and are taken from \cite{larson_electronic_2000}. Using these masses, the SEAQT electron $\tau^\epsilon$ values in the $xx$- and $zz$-directions as a function of the eigenenergies are reported in Figs. S5 and S6, respectively.
To evaluate the comprehensive thermoelectric characteristics of Bi$_2$Te$_3$,  Matthiessen's rule \cite{chen_nanoscale_2005} is used to derive an aggregate SEAQT electron relaxation parameter ($\tau^\epsilon$) value as a function of the eigenenergies given in Figs. S5 and S6. 

\begin{figure}
\centering
    \begin{minipage}{0.45\textwidth}
        \centering
        \includegraphics[width=8.5cm, height=6.5cm]{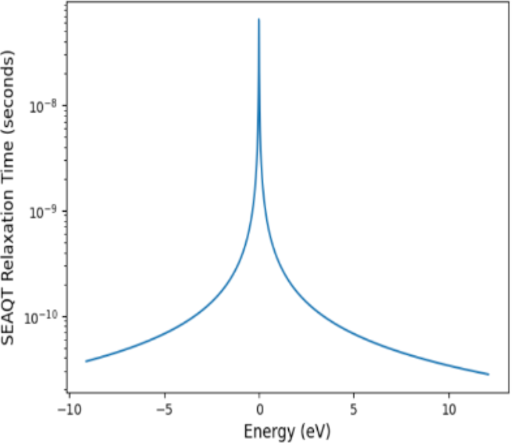} 
    \end{minipage}\hfill
\caption*{Figure S5: Calculated SEAQT electron relaxation parameter ($\tau ^\epsilon$) values for Bi$_2$Te$_3$ in the $xx$-direction as a function of the electron eigenenergies.}
\end{figure}

\begin{figure}
\centering
    \begin{minipage}{0.45\textwidth}
        \centering
        \includegraphics[width=8.5cm, height=6.5cm]{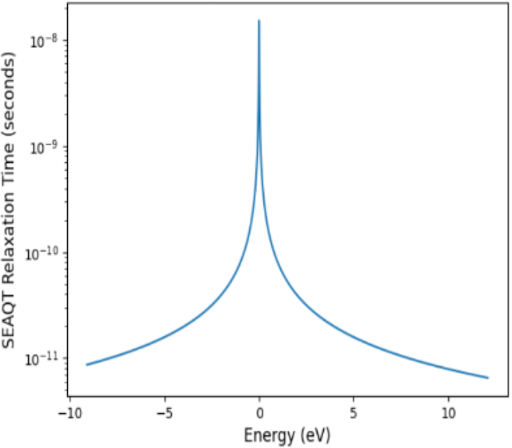} 
    \end{minipage}\hfill
\caption*{Figure S6: Calculated SEAQT electron relaxation parameter ($\tau ^\epsilon$ values
 for Bi$_2$Te$_3$ in the $zz$-direction as a function of the eigenenergies.}
\end{figure}

Fig. S7 shows that the SEAQT transport property predictions for Bi$_2$Te$_3$ relative to those from various experimental studies \cite{ge_enhanced_2018,li_thermoelectric_2012,ivanov_metal-ceramic_2022,yaprintsev_effects_2018} accurately capture the trends in the electrical and thermal conductivities, the Seebeck coefficient, and the figure of merit ($ZT$ factor). While the values align generally with the experimental data, disparities exist between the various experimental studies. For instance, the SEAQT thermal conductivity (Fig. S7b)) tends to be slightly higher compared to experimental data. However, this variance arises from predicted values based on a pristine crystal structure devoid of defects, whereas the experimental samples, fabricated via spark-plasma sintering, differ due to the presence of defects. The implications of this fabrication method are elaborated in Appendix C. Conversely, Fig. S8 compares the SEAQT outcomes for the phonon (lattice) thermal conductivity (i.e., excluding the electron contribution) with experimental \cite{satterthwaite_electrical_1957,goldsmid_thermal_1956} and molecular dynamic results \cite{qiu_molecular_2009}, showcasing significant agreement. Nevertheless, as evidenced in Fig. S8, excluding the electron contribution, notably underestimates total thermal conductivity for narrow-bandgap materials like Bi$_2$Te$_3$. Thus, simultaneous consideration of both electron and phonon contributions is imperative.

\begin{figure}
\centering
    a)
    \begin{minipage}{0.45\textwidth}
        \centering
        \includegraphics[width=8cm, height=5.5cm]{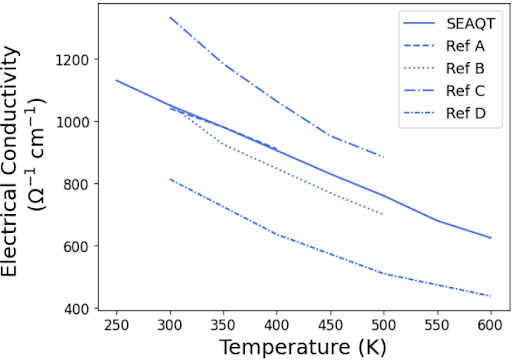} 
    \end{minipage}\hfill
    b)
    \begin{minipage}{0.45\textwidth}
        \centering
        \includegraphics[width=8cm, height=5.5cm]{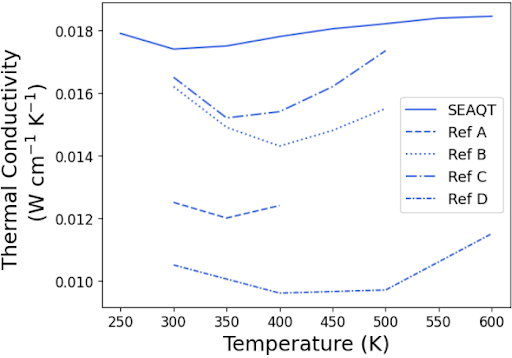} 
    \end{minipage}
\centering

    c)
    \begin{minipage}{0.45\textwidth}
        \centering
        \includegraphics[width=8cm, height=5.5cm]{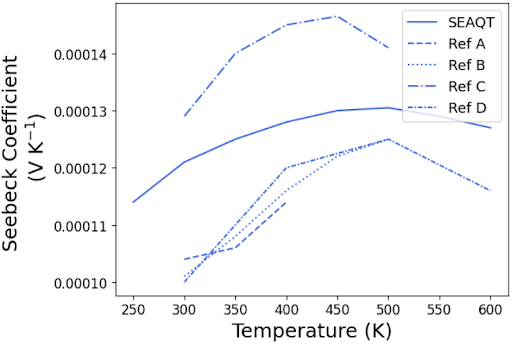} 
    \end{minipage}\hfill
    d)
    \begin{minipage}{0.45\textwidth}
        \centering
        \includegraphics[width=8cm, height=5.7cm]{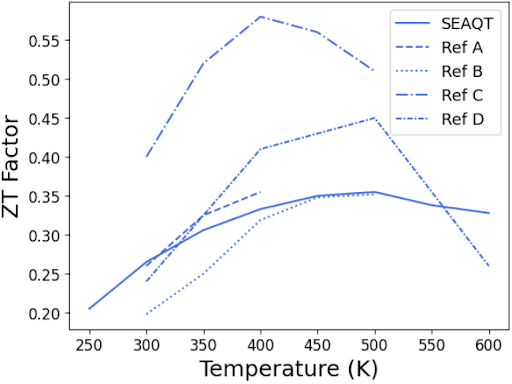} 
    \end{minipage}
\caption*{Figure S7: Comparison of the SEAQT results for the a) electrical conductivity, b) thermal conductivity, c) Seebeck coefficient, (d) and $ZT$ factor of Bi$_2$Te$_3$ with experimental data: Ref. A is \cite{ge_enhanced_2018}, Ref. B is \cite{li_thermoelectric_2012}, Ref. C is \cite{ivanov_metal-ceramic_2022}, Ref. D is \cite{yaprintsev_effects_2018}.} 
\end{figure}

\begin{figure}
\centering
    \begin{minipage}{0.45\textwidth}
        \centering
        \includegraphics[width=8.5cm, height=6.5cm]{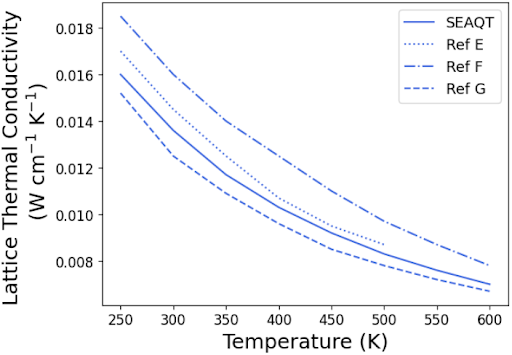} 
    \end{minipage}\hfill
\caption*{Figure S8: SEAQT Bi$_2$Te$_3$ phonon thermal (lattice) conductivity results compared with experimental (Ref F \cite{satterthwaite_electrical_1957} and Ref G \cite{goldsmid_thermal_1956}) and molecular dynamics results (Ref E \cite{qiu_molecular_2009}).}
\end{figure}

\section*{Appendix C: Bi$_2$Te$_3$ Thermal Conductivity Discussion}

Experimental Bi$_2$Te$_3$ samples typically exhibit higher levels of inherent defects, including line dislocations and grain boundaries, which reduce thermal conductivity. When separating the SEAQT Bi$_2$Te$_3$ thermal conductivity into lattice and electron contributions, the SEAQT phonon (lattice) thermal conductivity predictions closely match the experimental and molecular dynamic (MD) simulations as shown in Fig. S8, since the experimenal samples are relatively pure and the MD simulation conditions idealized. This is born out by comparing the experimental total thermal conductivity values in Fig. S7b), particularly at low temperatures, with those in Fig. S8. As can be seen, they are significantly lower than the experimental phonon thermal conductivity data in Fig. S8 due to phonon scattering on grain boundaries and dislocations \cite{ge_enhanced_2018}. These defects are intrinsic to the spark plasma sintering (SPS) fabrication method used for the Bi$_2$Te$_3$ samples in Fig. S7b) \cite{ge_enhanced_2018,li_thermoelectric_2012,ivanov_metal-ceramic_2022,yaprintsev_effects_2018}. Fabricating defect-free materials via SPS is challenging due to various mechanisms that need to be controlled, including temperature inhomogeneities, non-uniform stress distributions, die material, applied electric field, applied pressure, and heating rates \cite{anselmi-tamburini_spark_2021}. Despite ongoing research into controlling these mechanisms, primarily through plasma generation, the electroplastic effect, and Joule heating, SPS often results in materials with inherent defects that alter thermoelectric properties compared to single-crystal materials \cite{anselmi-tamburini_spark_2021,suarez_challenges_2013}. This issue is exacerbated by small grain sizes inherent to the SPS method due to the sintering of a micropowder, leading to significantly lower thermal conductivities compared to single-crystal samples \cite{suarez_challenges_2013}. While alternative methods have been shown to produce single-crystal Bi$_2$Te$_3$ samples with lower defect concentrations compared to SPS, there is a lack of thermoelectric property analysis within the temperature range of interest to compare with the SEAQT thermal conductivity data \cite{strieter_growth_1961,ainsworth_single_1956}.

\hfill \break

\hfill \break

\begin{table}[h!tbp]
\centering
\captionsetup{format=hang} \caption*{Table S1: Si and doped Si calculated SEAQT phonon relaxation parameter, $\tau^\epsilon$.}
\smallbreak
\begin{tabular}{| c | c |} 
  \hline
  Doping Concentration (1/cm$^{-3}$) & Calculated SEAQT Phonon Relaxation Parameter, $\tau^\epsilon$ (fs) \\ 
  \hline
  undoped   &  18.00\\
  \hline
  p-type 2 x 10$^{16}$ & 20.21 \\
  \hline
  p-type 1 x 10$^{20}$ & 41.04 \\
  \hline
  n-type 2 x 10$^{17}$ & 20.83 \\
  \hline
  n-type 1 x 10$^{18}$ & 33.47 \\
  \hline
  n-type 2 x 10$^{19}$ & 38.25 \\
  \hline
  n-type 1 x 10$^{20}$ & 90.72 \\
  \hline
\end{tabular}
\label{SupTable:1}
\end{table}

\begin{table}[h!tbp]
\centering
\captionsetup{format=hang} \caption*{Table S2: 
 Phonon group velocity and calculated SEAQT phonon relaxation parameter ($\tau^\epsilon$) values for the $xx$- and $zz$-directions of Bi$_2$Te$_3$.}
\smallbreak
\begin{tabular}{| c | c | c |} 
  \hline
  Direction & Phonon Velocity (m/s) & Calculated SEAQT Phonon Relaxation Parameter Values, $\tau^\epsilon (s)$ \\ 
  \hline
  $xx$ & 1800 & 1.4 x 10$^{-10}$ \\
  \hline
  $zz$ & 2600 & 6.7 x 10$^{-10}$\\
  \hline
\end{tabular}
\label{SupTable:2}
\end{table}

\begin{table}[h!tbp]
\centering
\captionsetup{format=hang} \caption*{Table S3: Bi$_2$Te$_3$ effective free electron masses for different directions.}
\smallbreak
\begin{tabular}{| c | c | c |} 
  \hline
  Direction & Effective Free Electron Mass & Effective Free Electron Mass  \\ 
   & for the Conduction  Band & for the Valance Band \\
  \hline
  $xx$ & 46.9 & 32.5 \\
  \hline
  $zz$ & 9.5 & 9.02\\
  \hline
\end{tabular}
\label{SupTable:3}
\end{table}

\hfill
\printbibliography
\emergencystretch 3em